\documentclass[epj]{svjour}
\usepackage{graphics}
\usepackage{epsfig}
\usepackage{amssymb}
\usepackage{amsmath}
\usepackage{bbm}
\begin{document}
\title{Decay, interference, and chaos:\\ How simple atoms mimic disorder}
\author{Andreas Krug\inst{1} \and Sandro Wimberger\inst{1,2} \and Andreas
Buchleitner\inst{1}% etc 
}                     % Do not remove
%
%\offprints{}          % Insert a name or remove this line
%
\institute{Max-Planck-Institut f{\"u}r Physik
  komplexer Systeme, N{\"o}thnitzer Str. 38, D-01187 Dresden \and 
Universit\`a degli Studi dell'Insubria, Via Valleggio 11, I-22100 Como} 
\date{Received: \today / Revised version: date}
% The correct dates will be entered by Springer
%
\abstract{
We establish a close quantitative analogy between the excitation and
ionization process of highly excited one electron Rydberg states under
microwave driving and charge transport across disordered 1D lattices. Our
results open a new arena for Anderson localization -- a disorder induced
effect -- in a large class of perfectly deterministic, decaying atomic systems.
 % end of PACS codes
} %end of abstract
\maketitle

\section{Introduction}
Probability transport in disordered media is an equally exciting and active
field at the very heart of statistical physics. Many particle dynamics in a
gas, turbulent hydrodynamic flow, the traffic flow across
canonically overcrowded european highways, and, last but not least, the 
flow of money across stock and option markets are described by statistical 
means. Also in the microscopic realm 
we are often confronted with similar situations where it is impossible to
track each single detail of the system under study -- either because we 
ignore the precise form of the potentials which generate the dynamics, or
because the number of individual constituents of the system is simply too
large, or a combination of both. Well-known examples thereof are the charge
transmission through mesoscopic wires, the scattering of
slow neutrons off heavy nuclei, or the ion transport across cellular
membranes. On this microscopic level, however, when noise is
sufficiently weak, quantum interference and tunneling can dramatically affect
classical probability transport, and weak \cite{cord} 
and strong (vulgo Anderson) 
localization \cite{anderson}, the Mott-Hubbard metal-insulator transition 
(see Zoller's contribution to this issue), 
Ericson fluctuations \cite{ericson}, 
chaos assisted tunneling \cite{ullmo}, or (universal) conductance
fluctuations \cite{lh94} 
are just the most prominent ones of the many surprising phenomena
which are born out by coherent complex transport on the microscopic scale. 

On a first glance, such signatures of complex dynamics (brought about by 
random potentials and/or many-particle interactions) appear precisely what 
we do {\em not} expect when dealing with well isolated atomic or quantum
optical systems. In the quest for an almost perfect control of matter, we 
seek optimal control of the 
potentials which determine the dynamics (see Gerber's contribution), 
in time and in space. Hence, how can
such low-dimensional systems 
mimic the complex dynamics of disordered systems? The answer is --
precisely -- Hamiltonian chaos, induced by nonlinear and perfectly
controlled coupling of the few (at least two are needed \cite{lichtenberg}) 
degrees of freedom
available. Given a sufficiently high density of states (in a quantum system
with a discrete or quasi-discrete spectrum), such coupling 
can destroy the symmetries, and this is the good
quantum numbers of the unperturbed
dynamics, giving rise to an extremely complex spectrum characterized by an
abundance of anticrossings of strongly variable size (see also 
Saenz' contribution) \cite{advances}. If we now prepare a wave 
function of
arbitrary shape at an arbitrary location in phase space, its time evolution
will indeed reflect the complex 
spectral structure, and, consequently, feature the characteristic properties
of complex/disordered systems. This is the essential consequence of the 
random matrix conjecture \cite{oriol}, 
which states
that the spectral properties of low dimensional quantum systems with
underlying chaotic classical dynamics exhibit the same (universal) 
statistical features as
complex quantum systems described by Random Matrix Theory (which took its
origin in the attempt \cite{wigner} 
to give a robust description of compound
nuclear reactions -- despite the little knowledge about the detailed nature of
the potentials which generate the experimentally observed cross sections). 

Which are the good experimental observables to monitor those features? 
In the usual setting of, e.g., charge transmission 
across a disordered
lattice,
transport occurs in configuration space, and the experimentalist 
measures a transmission probability. 
Such scenario can nowadays be mimicked in quantum optical
table top realizations of ``ideal'' lattices, where a suitable laser
configuration establishes a periodic (1D or 2D or 3D) potential, with
adjustable lattice constant and modulation depth. Cold atoms or ions
initially prepared at one specific location can then be monitored as they move
across the sample \cite{walther}, and it is equally 
possible to enforce the transition from 
localized to delocalized (in configuration space) eigenstates on the lattice
\cite{greiner}. 
On the other hand, we may ask with equal right whether certain
phenomena familiar from the theory of complex and/or disordered systems can be
imported to electronic dynamics on the scale of a single, simple atom, with
few degrees of freedom, and no {\em a priori} resemblence with familiar
transport problems. In other words, do disorder induced phenomena have any 
general and robust
bearing for the dynamics of simple atomic systems exposed to strong
perturbations? The answer is positive, and can be 
elaborated for various of the above-mentioned coherent transport 
phenomena. For reasons of time
and space, we shall specialize here to Anderson localization in driven atomic
systems, and, more specifically, in atomic Rydberg states under microwave
driving.

\section{Anderson localization on the atomic scale}

\begin{figure}
\centerline{\epsfig{figure=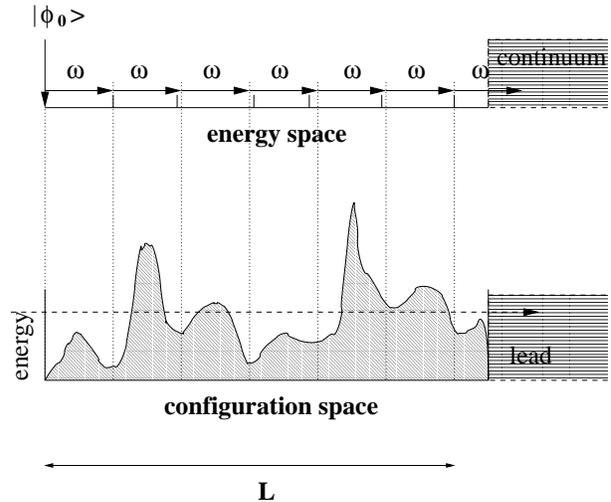,width=8cm}}
\caption{The Anderson scenario: A particle (dashed horizontal line) 
is transmitted across a disordered
potential in 1D configuration space. At each hump, the particle can be
reflected or transmitted, with random reflection/transmission coefficients. 
The atomic ionization problem under microwave driving at frequency $\omega$
is strictly analogous if we replace reflection and transmission by
emission and absorption of a photon into/from the driving field, and 
substitute configuration space by the energy axis. The ``atomic sample
length'' $L$ is then the ionization potential expressed in integer 
multiples of the photon
energy $\omega$, and the ``sample edge'' is defined by the continuum
threshold. The atomic decay to the continuum 
(indicated by the rightmost arrow across the ionization threshold) 
is mediated by the one photon transition matrix
element from the highest bound Rydberg state -- connected to the initial state
$|\phi _0\rangle$ (indicated by the vertical arrow) 
by a sequence of near resonant (though slightly detuned -- small verticle
ticks along the horizontal energy axis indicate the near resonantly coupled
bound states, dashed verticle lines highlight the lattice period) 
one photon transitions. The population of this last bound state depends
exponentially on $L/\xi$, with $\xi $ the characteristic decay length of the
localized wavefunction.}
\label{fig1}
\end{figure}

Anderson localization is a disorder induced effect, which was predicted 
\cite{anderson} by
Anderson to occur in the transmission of a charge across a one dimensional,
disordered lattice. As depicted in Fig.~\ref{fig1}, the problem cooks down to
the transmission of a particle across a one dimensional random potential, at
a given injection energy. The number of paths which guide the
particle from the left to the right of the sample is exponentially large,
since at each hump of the potential the particle can be either reflected or
transmitted (with randomly distributed transmission and reflection
coefficients), thus multiplying 
the available paths by a factor of two at 
each hump. The transmission amplitudes associated with these different paths
acquire randomly distributed phases as they migrate through the sample, and 
interfere {\em destructively} (if the phases get 
homogeneously distributed over the unit
circle) on exit at the right edge of the sample. Hence, a disordered potential
leads to the suppression of transmission by quantum interference, and a more
quantitative analysis shows that the eigenfunctions of the particle are
exponentially localized over the lattice, with a characteristic localization
length $\xi$. The final figure of merit which determines the measured
conductance across the sample is $\xi/L$, with $L$ the length of the 
sample.
$\xi/L $ determines the population of the last site right at the edge of the
sample, and hence the probability flux
which can escape from the sample, via the coupling matrix element connecting
the last site to the leads -- attached to the sample to probe the conductance. 

We now import this scenario to the atomic realm, more specifically to the
excitation and ionization dynamics of atomic Rydberg states under microwave
driving. As mentioned above, for features of complex/disordered transport to
become manifest on the atomic scale, we need a high density of states actively
involved in the dynamics. This
is guaranteed in the Rydberg regime under microwave driving, since the typical
Rydberg energy splitting lies in the microwave frequencies, and, consequently,
many Rydberg states will be efficiently coupled by the drive (through 
subsequent, near-resonant
one photon absorptions). 
The atomic transport process which we
want to parallel with the Anderson scenario is the transport of
electronic population initially prepared in a well defined atomic initial
state $|\phi_0\rangle =|n_0,\ \ell_0,\ m_0\rangle $ (where spherical quantum
numbers are used, as will be done throughout 
the sequal of this paper) towards the
atomic continuum -- easily measured as the ionization yield $P_{\rm ion}$ 
for given 
field amplitude $F$, field frequency $\omega $, and interaction time $t$. 
To make the analogy complete, we have to remind ourselves of the 
slow variation of the level splitting with energy in the Rydberg domain. 
As a consequence of  
this unharmonicity of the 
spectrum, the subsequent one photon transition matrix elements which establish
the ``hopping matrix elements'' between the different ``sites'' on the energy
axis mimic a
random series \cite{brenner96}, due 
to the imperfect matching (tantamount to quasi random 
detuning) of the driving frequency with the
exact transition frequency (much as the simplest random number generators
built on a modulo operation \cite{numrec}).
Consequently, if we replace ``transmission/reflection'' in the above
transmission problem \`a la Anderson by ``absorption/emission'' in the 
atomic excitation process, 1D configuration space by the energy axis, and the
attached leads by the atomic continuum,
we encounter precisely the same scenario: An
exponentially large number of paths visiting different sites (near resonantly
coupled Rydberg states in the atomic problem) accumulate quasi random phases
on exit from the sample, i.e., at the ionization threshold. 
If one photon transitions from the last near resonantly
coupled bound state to the continuum define the dominant ionization channel,
the atomic ionization yield will be proportional to this state's population
-- determined by the localization parameter
$\xi /L$ already familiar from the Anderson scenario. Again, the sample
length $L$ is given by the number of sites which span the lattice, equivalent
to 
the ionization potential of the atomic initial state in multiples of the
driving field photon energy. Provided $\xi /L\ll 1$, the 
Anderson model predicts
{\em exponential} suppression of the ionization yield due to quantum
interference. Despite the strong, near resonant driving, the atom will 
cease absorbing energy from the field! 

In fact, there already are abundant experimental data 
\cite{galvez88,bayfield89,arndt91,noel2000} which indicate that 
the above
mechanism is at work in driven Rydberg states of atomic hydrogen and of alkali
atoms. The central experimental result is an increase of the ``scaled
ionization threshold amplitude'' $F_0^{(10\%)}=F^{(10\%)}n_0^4$ with
increasing principal quantum number $n_0$ of the initial atomic state. 
In other words, the field
amplitude $F_0^{(10\%)}$ 
required to ionize $10\%$ of the atoms, at given field frequency
$\omega$ and interaction time $t$, measured in units of the average 
Coulomb field
$\sim n_0^{-4}$ experienced by the electron on its unperturbed initial Kepler
orbit, {\em increases} as we {\em decrease} the ionization potential 
of the atomic initial
state! This is in dramatic contradiction with the result of a classical
treatment, which predicts systematically smaller ionization thresholds (due to
classically chaotic phase space transport) than observed in the
experiment. Consequently, there exists a range of field amplitudes which
induce efficient classical ionization, whilst the experimentally observed
yield is close to zero. This quantum suppression of classically chaotic
ionization is interpreted as a signature of Anderson localization, and
traditionally termed ``dynamical localization'', such as to identify dynamical
chaos as the cause of the quasi-ran\-do\-mi\-za\-tion of 
the hopping matrix elements
in the Anderson picture. 
However, 
this experimental finding is only consistent with the hypothesis of
Anderson localization, it is not a clear proof -- simply since one may imagine
different mechanisms which inhibit the ionization process, such as 
semiclassical
stabilization effects caused by (partial) barriers in phase space
\cite{graham,benson95,physrep}. Furthermore,
most experiments have been performed under slightly different experimental
conditions: not only have different atomic species been used, but also
different atomic initial states were exposed to microwave fields of different
frequencies $\omega $ and of variable duration $t$. All available
experimental data exhibit large quantitative differences between the
ionization threshold of atomic hydrogen on the one hand and of nonhydrogenic
initial states of alkali atoms on the other, with the alkali thresholds down
by a factor five to ten as compared to the hydrogen thresholds
\cite{benson95}. 
The original
theory of dynamical localization in driven atomic systems
\cite{fishman82}, based on a very
crude model of the actual bound state 
atomic dynamics, is not suited to explain these quantitative
differences nor to definitively exclude processes distinct from Anderson 
localization which might cause the observed increase of $F_0^{(10\%)}$.

What we therefore need is an accurate theoretical treatment of microwave
driven one electron Rydberg states of atomic hydrogen and of alkali atoms.
Such treatment is not only required
to explain the experimental findings so far available in a consistent 
and unified way, but also 
to guide future experiments which seek to test quantitative predictions
which follow from Anderson localization theory for the atomic ionization
process. 

\section{An accurate treatment of one electron Rydberg states under
electromagnetic driving}

Our specific problem is described by the $2\pi/\omega $-periodic 
Hamiltonian
\begin{equation}
H(t)={\frac {{\bf p}^2} {2}}+V_{\rm atom}(r)+{\bf F} \cdot {\bf r } \cos 
\omega t,\  r>0,
\label{hamil}
\end{equation}
where $V_{\rm atom}(r)$ denotes the atomic potential seen by the valence 
electron, which is taken care of by a R-matrix approach
\cite{krug}.\footnote{The 
R-matrix treatment requires input of the quantum defects 
of the non-hydrogenic angular momentum states of the unperturbed alkali
atoms, known from high precision spectroscopy experiments \cite{lorenzen}.}
The periodicity of $H$ suggests to explore Floquet 
theory \cite{floquet,buchl95}, and we therefore 
end up solving the Floquet eigenvalue problem 
\begin{equation}
{\cal H} |\varepsilon _j\rangle =\varepsilon _j |\varepsilon _j\rangle \, ,
\label{floqew}
\end{equation}
with the Floquet Hamiltonian
\begin{equation}
{\cal H}=H-{\rm i}{\partial_t},
\label{ewgl}
\end{equation} 
the spectrum of which is 
invariant under translations by
$\omega$.
Knowledge of the $|\varepsilon_j\rangle $ and $\varepsilon_j$ 
within an energy range of width $\omega$
is therefore 
sufficient for a complete description of the dynamics.

After a further Fourier transform of the $2\pi/\omega $-periodic
$|\varepsilon_j\rangle $, amended by complex dilation of the 
Hamiltonian \cite{buchl95,combes}, the eigenvalue problem is represented in a
real 
Sturmian basis, what converts (\ref{ewgl}) into a block-tridiagonal, 
complex symmetric, sparse banded eigenvalue problem
\cite{krug,buchl95,wirwannwo}.  
Note that the strong selection rules induced by the Sturmian basis are
absolutely crucial for the numerical treatment of our problem,
since they allow for a considerable reduction of the required memory.
In the parameter range 
typically employed in the laboratory, typical dimensions are 
$10^6\times 10^4$, what nontheless requires a very large parallel 
supercomputer. Indeed, all the results presented hereafter were
obtained on the CRAY T3E of the High Performance Computing Center RZG of the
Max-Planck-Society at Garching and, once this machine 
became too small, on 
the Hitachi SR8000-F1 of the Bavarian Academy of Sciences in Munich.
Though, the availability of such a large machine is not enough.
In addition, an efficient parallel implementation of
the Lanczos diagonalization routine is needed, which is by no means a trivial
requirement, due to considerable communication between different (and not only
adjacent) processors of the parallel machine. 

Once the theoretical and numerical machinery briefly sketched above is 
assembled, we can start our ``numerical experiment'', which closely mimics
the reality in the laboratory. The ``counts'' which the
numerical experimentalist has to collect much as his colleague in the real lab
are the poles of the resolvent operator in the lower half of 
the complex plane \cite{akab02}, i.e. the  complex eigenvalues $\varepsilon _j
=E_j-i\Gamma_j/2$ of the above eigenvalue
problem. 
From these we can immediately extract the ionization yield \cite{buchl95}
\begin{equation}
P_{\rm ion}=1-\sum_{\varepsilon}|\langle\phi_0|\varepsilon\rangle |^2\exp
(-\Gamma_{\varepsilon}t)\, ,
\label{sig}
\end{equation}
for the specific choice of $|\phi_0\rangle $, $t$, $\omega $, and $F$.
Note that approx. $4000\ldots 10000$ poles are typically collected here. 
This is simply due to the fact that the decomposition of
the atomic initial state $|\phi_0\rangle $ in the dressed state basis is
extremely broad in our present case \cite{swakab}, what is a direct 
consequence of the
efficient destruction of good quantum numbers by the driving microwave 
field, and,
hence, of {\em quantum chaos}. 

\section{A unified view on Anderson localization in driven atoms}

\begin{figure}
\centerline{\epsfig{figure=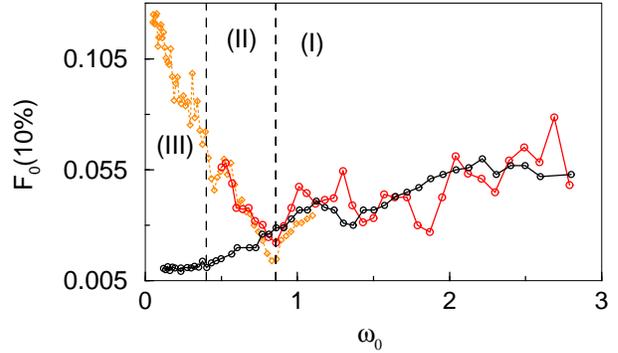,width=8cm}}
\caption{Comparison of the ionization threshold of atomic hydrogen 
(obtained in laboratory experiments \protect\cite{galvez88,pmk}; 
light
symbols) and of lithium ($\ell _0=0$) Rydberg states 
(from our numerical experiment;
dark symbols) under microwave driving of duration $t=327\times 2\pi/\omega$
at $\omega/2\pi =36\ \rm GHz$, in the range $n_0=28\ldots 80$ of principal
quantum numbers (from left to right, since $\omega_0=\omega n_0^3$). Since the
laboratory values of the physical parameters are identical, the present plot
in scaled units does not imply any a priori assumptions on alkali scaling
laws!}  
\label{fig4}
\end{figure}

With the above, we can now perform a direct comparison of the ionization
thresholds of atomic Rydberg states of hydrogen and of those of alkali Rydberg
states, under {\em precisely equivalent} conditions. 
For the sake of comparison
to the arguably largest experimental data set (produced in the Stony Brook 
group \cite{pmk}), we specifically 
choose $\omega/2\pi =36\ \rm GHz$, $t=327\times
2\pi/\omega $, and $n_0=28\ldots 80$, $\ell_0=m_0=0$, where $m_0$ is conserved
under linearly polarized driving. Precisely as in the laboratory, we 
scan $F$ from low to high values, monitor the dependence of $P_{\rm ion}$ on
$F$, and extract $F_0^{(10\%)}$, for increasing values of
$n_0$. Fig.~\ref{fig4} shows a comparison of our numerical results for lithium 
$\ell_0=0$ states to laboratory results on atomic hydrogen. Surprisingly, and
{\em in perfect contradiction} to all published experimental evidence, the
nonhydrogenic (the $\ell_0=0$ state carries the largest quantum defect
$\delta_{\ell_0=0}=0.399468$) lithium Rydberg states exhibit essentially {\em
the same} ionization thresholds as atomic hydrogen in regime (I), i.e. for
``scaled frequencies'' $\omega_0=\omega n_0^3 > 0.8$. 
Only for decreasing values 
of $n_0$ leading
to smaller values of $\omega_0$ do we observe an increasing
discrepancy between the hydrogenic and alkali thresholds. It turns our that
this can be traced back to the local density of states in the Rydberg series
of the different atomic species \cite{wirwannwo}: 
For $\omega $ larger than the local {\em
hydrogenic} level spacing -- which scales like $n_0^{-3}$, hence 
$\omega_0 \gtrsim 1$ -- the atomic excitation process is essentially unaffected by
the spectral substructure in the alkalis. The atomic system offers a ladder of
near resonant one photon transitions connecting $|\phi_0\rangle $ to the
continuum (see Fig.~\ref{fig1}), and it is irrelevant for the transport 
process whether this occurs
under the participation or in the absence of non-hydrogenic states.
However, once the driving frequency is smaller than the hydrogenic level
spacing but still larger than the splitting between the non-hydrogenic alkali
initial state and the hydrogenic manifold, the same sequence of near resonant
one photon transitions can still be established in the alkali atom, whilst in
hydrogen only sequences of higher order transitions are left to mediate the
ionization process. Hence, the realm of the Anderson scenario outlined in the
introduction extends over a wider range of principal quantum numbers in the
alkali species than in atomic hydrogen, and this is indeed manifest in
Fig.~\ref{fig4}: In regime (II), with $0.4 \lesssim \omega_0 \lesssim 0.8$, 
the lithium data
continue to exhibit decreasing thresholds with decreasing $\omega _0$, as an
indicator of Anderson localization, whilst the hydrogenic thesholds rapidly
increase as the binding potential is increased (and the local density of
states is decreased, as compared to the photon energy). 
Finally, in regime (III), also the local energy splitting in the alkali atom
is too large to be driven near resonantly by a single photon. The Anderson
scenario breaks down, and the thresholds saturate in the plot of
Fig.~\ref{fig4}. 

The above results hold, at least, a two-fold message: First, there is the
prediction that comparable
ionization thresholds for alkali and hydrogen Rydberg states will be observed 
once the driving
frequency is larger than the spacing between adjacent hydrogenic manifolds. 
This
quantitative prediciton can be checked immediately in state of the art
laboratory experiments \cite{noel2000}. Second, when re-analyzing all 
available experimental 
data
on the microwave ionization of non-hydrogenic alkali Rydberg states by scaling
the driving field amplitude and frequency 
as done in Fig.~\ref{fig4}, we find \cite{Diss} that all
these data were obtained in regimes (II) and (III), what explains the
apparently dramaticly enhanced ionization yields of non-hydrogenic alkali
Rydberg states as compared to hydrogen \cite{wirwannwo,Diss}. 
Note that this discrepancy remained a
puzzle over more than one decade \cite{pmk}, since, so far, precisely identical
experimental conditions for different atomic species 
were never established in the lab, and the appropriate
scaling of $F$ and $\omega $ was controversial due to the badly defined
classical analog of the alkali Rydberg dynamics \cite{arndt91,benson95,pmk}. 
As a matter of fact, our
comparison of hydrogen and lithium data in terms of scaled units in
Fig.~\ref{fig4} do {\em not} imply any a priori  assumptions, since they were 
obtained for the same laboratory values of all physical parameters, but the
fact that alkali and hydrogen thresholds coincide in regime (I) {\em proves} 
that, for sufficiently high driving
frequencies,
the hydrogenic scaling holds even for alkalis! 
Even more, we do know by now \cite{Diss} 
that also non-hydrogenic initial
states of sodium and rubidium
exhibit the same, hydrogenic, threshold in regime (I), what strongly suggests
that this is a {\em universal threshold for one-electron Rydberg 
states}. Only this universality makes the increase of $F_0^{(10\%)}$ with
$\omega _0$ or $n_0^3$ a sufficient condition for the Anderson scenario to
prevail, since only the universality shows 
that the {\em only} relevant ingredient is the local density
of states compared to the photon energy.

\begin{figure}
\centerline{\epsfig{figure=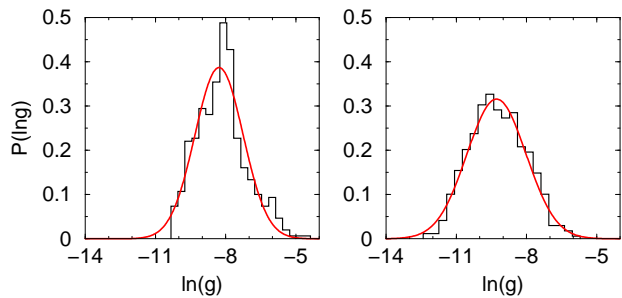,width=4cm,angle=-90}}
\caption{Distribution (histograms) 
of the atomic conductance $g$ of a 1D Rydberg atom
\protect\cite{abigjz98,sawda}, sampled over $500$ different realizations of
the localization parameter $\xi/L=0.2$, in the frequency range
$\omega_0=2.0\ldots 2.5$, for $n_0=40$ (left) and $n_0=100$ (right). The
lognormal fit (smooth curve) is excellent for $n_0=100$, in perfect
quantitative agreement
with Anderson localization theory \protect\cite{pichard}. The deviations from
lognormal behaviour for $n_0=40$ reflect a finite size effect.} 
\label{fig5}
\end{figure}

There is another consequence of Anderson localization
theory which can be imported to our atomic system: The localization length
$\xi$ is really well-defined a quantity only in the limit of infinite sample
length, and will fluctuate around this limiting value for finite $L$
\cite{pichard}. The
theoretical modelling of the conductance across an Anderson localized 1D wire
of finite length $L$ predicts a normal distribution of the logarithm of the 
properly normalized conductance $g$, if sampled over different realizations of
the random lattice potential (see Fig.~\ref{fig1}), {\em at fixed} $\xi/L$
\cite{pichard}. 
Since the (mesoscopic) conductance is given by nothing but a sum over
transition matrix elements (from left to right of the sample) \cite{landauer}, 
we can come up
with an analogous definition of the ``atomic conductance'' in our ionization
problem \cite{abigjz98},
\begin{equation}
g:=\frac{1}{\Delta}\sum_j|\langle\phi_0|\varepsilon _j\rangle|^2
\Gamma_j\,  ,
\label{condu}
\end{equation}
since the decay rates $\Gamma_j$ can be understood as transition
matrix elements of a suitably defined Floquet scattering problem
\cite{sawda} (with $\Delta $ the average level spacing). 
Now, if 
Anderson is at work in the atomic problem, the atomic conductance has to 
exhibit the same statistical properties as required by the theory of
disordered transport -- 
and, indeed, it does! Fig.~\ref{fig5} shows the distribution of the 
atomic conductance of a 1D model atom \cite{abigjz98} (the electron 
is confined to the 
configuration space axis parallel to the polarization of the driving field),
for fixed $\xi/L=0.2$ (which is given in terms of $n_0,F,\omega$, 
according to the
original theory of dynamical localization \cite{fishman82}),
and for two different values
$n_0=40,100$ of the principal quantum number.\footnote{For each
plot 
approx. 500 complex-valued spectra have been sampled, making such statistics
even for the 1D model atom a rather expensive enterprise. By now, however, we
have first evidence from 3D calculations that the same result can be expected
for the real atom \cite{swakab}.} The lognormal fit to the data is
{\em excellent} for $n_0=100$, what represents another, independent and 
{\em quantitative} indicator of Anderson localization as the dominant
mechanism which determines the energy exchange between the atom and the
field. We also observe in Fig.~\ref{fig5} that the fit is significantly worse
for $n_0=40$, what, however, does not contradict our preceding statement: At
too low quantum numbers, the atomic sample size $L$, i.e. the number of near
resonantly coupled Rydberg states between $|\phi_0\rangle $ and the continuum
threshold is too small ($L\simeq 10$) to allow for a smooth exponential
localization of the electronic wavefunction over the energy axis ($\xi
=0.2L=2$).  The figure
therefore simply highlights a finite size effect. 

\begin{figure}
\centerline{\epsfig{figure=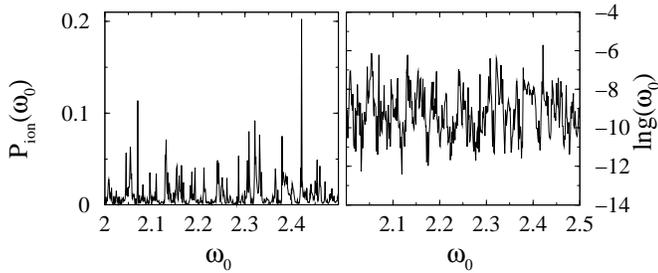,width=4cm,angle=-90}}
\caption{Ionization yield $P_{\rm ion}$ (left) and atomic conductance $g$ 
(right) of a 1D
Rydberg atom \protect\cite{abigjz98,sawda}, as a function
of the scaled frequency $\omega _0$, in the Anderson/dynamically 
localized regime at 
 $\xi/L=0.2$, $n_0=100$ (below threshold in Fig.~\ref{fig4}).} 
\label{fig6}
\end{figure}

Finally, Fig.~\ref{fig6} shows the physical imprint of the lognormal
distribution of Fig.~\ref{fig5}, the dependence of the ionization yield
$P_{\rm ion}$ on the scaled frequency $\omega _0$, at 
$\xi/L=0.2$ (below threshold in Fig.~\ref{fig4}): The yield is typically very
small, close to zero, but exhibits strong, erratic enhancements at particular
values of $\omega _0$. Correspondingly, the atomic conductance $g$ fluctuates 
{\em on a logarithmic scale}, in the right column of the figure. This is
nothing but the immediate signature of the highly sensitive interference of
the many paths defined by subsequent absorption and emission events mediating
the transition from the initial state to the atomic continuum (see
Fig.~\ref{fig1}), and a specific signature of Anderson
localization. The reader should contemplate this strongly fluctuating
signal, since what fluctuates is a quantity obtained from a
{\em weighted average} over the entire Floquet spectrum, and not just 
one single 
rate $\Gamma_j$!

\section{Conclusion}

We provided two independent, quantitative proofs for Anderson
localization as the dominant mechanism which governs the excitation and
ionization process of strongly driven one electron Rydberg systems. Given the
universal ionization threshold which we observed for atomic hydrogen and
lithium Rydberg states in the high frequency parameter range, it appears
legitimate to speculate that this disorder-induced quantum interference effect
can be generalized for an even larger class of driven atomic or molecular
system, {\em if only} the simple Anderson scenario can be etablished {\em on
the energy axis}, as depicted in Fig.~\ref{fig1}. 

\section{Acknowledgements}
This paper is dedicated to Peter Lambropoulos, and might serve to
illustrate a remark
due to Bertrand Russell -
\begin{quote}
\raggedright
{\em The pursuit of quantitative precision is as arduous as it is important.}
\end{quote}
in his ``ABC of relativity'', which could be due to Peter 
as well.
 
Support as 
a Grand Challenge project at the Leibniz-Rechenzentrum of the Bavarian Academy
of Sciences is most gratefully acknowledged.

\end{document}